\documentclass[letterpaper,preprintnumbers,prd,twocolumn,nofootinbib,nobibnotes,showpacs]{revtex4}
\usepackage{epsfig}
\usepackage{graphicx}%
\usepackage{dcolumn}
\usepackage{amsmath}

\makeatletter
\def\btt#1{\texttt{\@backslashchar#1}}%
\DeclareRobustCommand\bblash{\btt{\@backslashchar}}%
\makeatother

\begin{document}

\title{A Universe Dominated by Dilaton Field}

\author{Chang Jun Gao$^1$}\email{gaocj@mail.tsinghua.edu.cn}\author{Shuang Nan Zhang$^{1,2,3,4}$}
\email{zhangsn@mail.tsinghua.edu.cn}\affiliation{$^1$Department of
Physics and Center for Astrophysics, Tsinghua University, Beijing
100084, China(mailaddress)} \affiliation{$^2$Physics Department,
University of Alabama in Huntsville, AL 35899, USA}
\affiliation{$^3$Space Science Laboratory, NASA Marshall Space
Flight Center, SD50, Huntsville, AL 35812, USA}
\affiliation{$^4$Key Laboratory of Particle Astrophysics,
Institute of High Energy Physics, Chinese Academy of Sciences,
Beijing 100039, China}

\begin{abstract}
Using a single dilaton field, a unified model of the Universe is
proposed, which evolves from the radiation-like dominance in the Big
Bang, to the dark-matter-like dominance in the early Universe, to the
coexistence of both dark-matter-like and dark energy
today, and finally to the dark energy dominance in the infinite
future. This model is consistent with current results on the age
of the Universe, the transition redshift from deceleration to
acceleration, BBN and evolution of dark energy. Future higher
quality data may constrain the cosmic evolution of dark matter, dark
energy and Hubble constant more precisely and make critical tests
on our model predictions.
\end{abstract}

\pacs{98.80.Cq, 98.65.Dx}
\maketitle
\section{Introduction}
The observations of supernova [1, 2], cosmic microwave
background [3, 4] and large scale structure [5, 6] in recent years
indicate that the Universe is accelerating and thus some form of
dark energy must exist in the Universe to drive this acceleration.
Investigation on the nature of dark energy becomes one of the most
important tasks for modern physics and modern astrophysics. Up to
now, many candidates of dark energy have been proposed to fit
various observations which include the Einstein's cosmological
constant [7], quintessence [8], k-essence [9], tachyon [10],
phantoms [11], brans [12] and so on. There have also been some
models of unified dark matter and dark energy [13]. Padmanabhan
and Choudhury developed a model of unified matter and dark energy
using a single tachyon field to explain both clustered dark matter
at small scales and smooth dark energy at large scales.
Gonzalez-Diaz investigated a model of the Universe filled with a
generalized Chaplygin fluid which mimics both dark matter and dark
energy. Cardone et al have used a phenomenological approach to
study the problem of unified dark matter and dark energy.
Susperregi has also developed a cosmological scenario where the
dark matter and dark energy are two simultaneous manifestations of
an inhomogeneous dilaton field. In his study, Susperregi has
constructed a dilaton potential with the ``trough'' feature.\\
\hspace*{3mm} In this paper, we present a model of the Universe
dominated by the dilaton field with a Liouville type potential.
The potential is the counterpart of the Einstein's cosmological
constant in the dilaton gravity theory. Since it can be reduced to
the Einstein cosmological constant when the dilaton field is set
to
zero, we call it the cosmological constant term in the dilaton gravity theory. \\
\section{dynamic equations}
In Ref.[14], we have derived the action of dilaton field in the
presence of Einstein's cosmological constant
\begin{eqnarray}
S=\int
d^4x\sqrt{-g}\left[R-2\partial_{\mu}\phi\partial^{\mu}\phi-V\left(\phi\right)\right],
\end{eqnarray}
where
\begin{eqnarray}
V\left(\phi\right)&=&\frac{2\lambda}{3\left(1+\alpha^2\right)^2}\left[\alpha^2\left(3\alpha^2-1\right)e^{-2\phi/\alpha}
\right.\nonumber\\&&\left.+\left(3-\alpha^2\right)e^{2\phi\alpha}+8\alpha^2e^{\phi\alpha-\phi/\alpha}\right],
\end{eqnarray}
is the Liouville-type potential with respect to the cosmological
constant, $\phi$ is the dilaton field, $\lambda$ is the Einstein's
cosmological constant and $\alpha$ is a free parameter which
governs the strength of the coupling of the dilaton to the
Einstein's cosmological constant. When $\alpha=0$ or $\phi=0$, the
action reduces to the usual Einstein scalar theory and the
potential becomes a pure cosmological constant. Here the usual
Einstein's cosmological constant term appears not as a constant
but as a coupling to the dilaton field which reveals the
interaction of vacuum energy (i.e. the dark energy today) and
dilaton matter. This implies that the potential plays the role of
both matter and dark energy. In other words, matter and dark
energy might both originate from this unique potential. This is
the starting point of our discussion. In the following we will
investigate whether this potential can mimic the total energy of
our Universe. When $\alpha=\pm\sqrt{1/3},\pm1,\pm\sqrt{3}$, the
action is just the SUSY potential in sting theory. It is apparent
that changing the sign of $\alpha$ is equivalent to changing the
sign of $\phi$.
Thus it is sufficient to consider only $\alpha > 0$ while $\phi$ may be positive or negative.\\
\hspace*{3mm}Recent measurements of the power spectrum of the
cosmic microwave background detected a sharp peak around $l\approx
200$, indicating that the Universe is highly likely flat. Thus we
only consider the flat Universe which is dominated by the
spatially homogeneous dilaton field and
described by the flat Friedmann-Robertson-Walker metric.\\
\hspace*{3mm}The equations of motion can be reduced to three
equations
\begin{eqnarray}
\frac{\dot{a}^2}{a^2}+\frac{2\ddot{a}}{a}&=&-\dot{\phi}^2+\frac{1}{2}V,\ \ \ \ 3\frac{\dot{a}^2}{a^2}=\dot{\phi}^2+\frac{1}{2}V,\nonumber\\
\ddot{\phi}+3\frac{\dot{a}}{a}\dot{\phi}&=&-\frac{1}{4}\frac{\partial
V }{\partial \phi},
\end{eqnarray}
where dot denotes the derivative with respect to $t$ and $a(t)$ is
the scale factor of the Universe. We find that the last equation
can be derived from the former two. So it is sufficient to solve
the first and the second equation. It follows immediately that
\begin{equation}
3H^2=\left(\frac{{dH}}{d\phi}\right)^2+\frac{1}{2}V,
\end{equation}
where $H\left[\phi\left(t\right)\right]\equiv{\dot{a}}/{a}$ is the Hubble parameter.\\
\hspace*{3mm}For an arbitrary $\alpha$, a solution of Eq.(4) is
given by
\begin{equation}
H=\frac{1}{1+\alpha^2}\sqrt{\frac{\lambda}{3}}\left(e^{\phi\alpha}+\alpha^2
e^{-\phi/\alpha}\right).
\end{equation}
Since there is no integration constant in the solution $H(\phi)$,
it is apparent $H(\phi)$ is a special solution to Eq.(4). We are
interested in that this special solution can actually describe the
evolution of our Universe. After inserting $H(\phi)$ into Eqs.(3),
we can obtain the expressions of $a(\phi)$ and $\phi(t)$. So the equations of motion are resolved.\\
\section{mimic our universe}
We can now mimic the total energy of the Universe using the single
dilaton field. Taking into account of $\alpha>0$, we find from
$dH/d\phi=-\dot{\phi}$ that when $\phi> 0$, $\phi$ is the
monotonically decreasing function of $t\in \left(-\infty,
+\infty\right)$. When $\phi < 0$, $\phi$ is the monotonically
increasing function of $t\in \left(-\infty, +\infty\right)$. Thus
the positive dilaton field can never evolve to the negative one
and vice versa. Without the loss of generality, we consider only
the case of $\phi>0$.\\
\hspace*{3mm}From the equations of motion we can derive the scale
factor of the Universe $a(\phi)$, the energy density of dilaton
field $\rho$, the age of the present Universe $\tau$ and the
parameter of equation of state $w=p/\rho$,
\begin{eqnarray}
a&=&e^{-\int_{\phi_0}^{\phi}H\left(dH/d\phi\right)^{-1}d\phi}, \ \
\ \ \rho=\frac{3}{8\pi}H^2, \ \ \
\nonumber\\
\tau&=&-\int_{\phi_i}^{\phi_{0}}\left(\frac{dH}{d\phi}\right)^{-1}d\phi,\
\ \ w=1-\frac{V}{8\pi\rho},
\end{eqnarray}
where $\phi_i$ is the value of the field at the time of $t=t_p$
($t_p$ is the Planck time) and $\phi_0$ is the value today. Since
the evolution of the Universe is from $\phi_i$ to $\phi_0$, so
$\phi_i$ is the lower limit and $\phi_0$ is the upper one. We have
set the scale factor today equals to $1$. \\
\hspace*{3mm}Among the equations of motion only two are independent, we can now recall
the specific initial conditions with respect to $\rho(\phi)$ and $w(\phi)$ in Eqs.(6).
Throughout the paper we will employ the Planck units in which the speed of light $c$,
gravitational constant $G$ and Planck constant $\hbar$ are all set equal to $1$. Then
we have the Planck energy density $\rho_p=1$, Planck time $t_p=1$. Since our discussion
is starting from the Planck time when the Universe was filled with the extremely
relativistic particles (equation of state $p=1/3\rho$) as required by the standard Big
Bang model, so we have the energy density $\rho_i=1$ and state equation $w_i=1/3$ at
the time of $t=t_p$. Given any value of $\alpha$ in the expression of $w$, we can
numerically calculate the relationship between the equation of state $w$ and $\phi$, as
shown in Fig.1. We find that when $\alpha=1.414213562\simeq \sqrt{2}$ and
$\phi\rightarrow\infty\Longleftrightarrow z\rightarrow\infty$, $w\rightarrow 1/3$,
which corresponds to the equation of state of the Universe at the Planck time.
Consequently we take $\alpha= \sqrt{2}$ in the rest of the paper.\\
\hspace*{3mm}For the present universe, it has the critical energy
density (to make it flat) and that it consists of
$\Omega_{m0}=0.30\pm0.04$ [15] of matter and
$\Omega_{x0}=0.70\pm0.04$ [15] of dark energy. Then from the
current parameters of equation of state, for matter $w_{m0}=0$ and
dark energy $w_{x0}=-1.02_{-0.19}^{+0.13}$ [16], we get the
parameter of the equation of state for the total energy
$w_0=w_{x0}\Omega_{x0}/\left(\Omega_{m0}+\Omega_{x0}\right)$; the
critical energy density $\rho_0=3.6h^2\times 10^{-123}$, where $h$
is in the range of $0.70_{-0.03}^{+0.04}$ [15] for the present
Universe. In Table $\textrm{I}$, we summarize the age of the
Universe $\tau$ and the transition redshift from deceleration to
acceleration $z_T$ evaluated with different combinations of
$\Omega_{m0}$, $w_{x0}$ and
$h$ within their ranges given above. For brevity, we only show these certain combinations whose predictions are consistent with current results on the age of the Universe and the transition redshift.\\
\begin{table}
\caption{Cosmic parameters evaluated with different $\Omega_{m0}$,
$w_{x0}$ and $h$. The dimensions: $\tau(10^{10} yr)$.
\label{table:J_average}}
\begin{tabular}{cccccccccccc}
\hline\hline $\textrm{Initial}$ & ${}$ & $\textrm{Conditions}$ &
${\|}$ & \ \ \ \ $ \textrm{Results}$ \\ \hline\hline
$\Omega_{m0}$ & $w_{x0}$ & $h$  & ${\|}$ & $\tau$ & $z_T$\\
\hline
$0.26$ & $-0.92$ & $0.67$  & ${\|}$ & $1.32$ & $0.52$   \\
\hline
$0.26$ & $-0.92$ & $0.70$  & $\|$ & $1.26$ & $0.52$  \\
\hline
$0.26$ & $-0.92$ & $0.74$  & $\|$ & $1.19$ & $0.52$   \\
\hline
$0.26$ & $-1.00$ & $0.67$  & $\|$ & $1.41$ & $0.68$   \\
\hline
$0.26$ & $-1.00$ & $0.70$  & $\|$ & $1.35$ & $0.68$  \\
\hline
$0.26$ & $-1.00$ & $0.74$  & $\|$ & $1.28$ & $0.68$ \\
\hline
$0.26$ & $-1.055$ & $0.67$  & $\|$ & $1.49$ & $0.82$  \\
\hline
$0.26$ & $-1.055$ & $0.70$  & $\|$ & $1.43$ & $0.82$   \\
\hline
$0.26$ & $-1.055$ & $0.74$  & $\|$ & $1.35$ & $0.82$  \\
\hline\hline
$0.30$ & $-0.97$ & $0.67$  & $\|$ & $1.31$ & $0.52$   \\
\hline
$0.30$ & $-0.97$ & $0.70$  & $\|$ & $1.26$ & $0.52$   \\
\hline
$0.30$ & $-0.97$ & $0.74$  & $\|$ & $1.19$ & $0.52$   \\
\hline
$0.30$ & $-1.00$ & $0.67$  & $\|$ & $1.34$ & $0.57$  \\
\hline
$0.30$ & $-1.00$ & $0.70$  & $\|$ & $1.28$ & $0.57$   \\
\hline
$0.30$ & $-1.00$ & $0.74$  & $\|$ & $1.22$ & $0.57$ \\
\hline
$0.30$ & $-1.115$ & $0.67$  & $\|$ & $1.49$ & $0.82$   \\
\hline
$0.30$ & $-1.115$ & $0.70$  & $\|$ & $1.43$ & $0.82$   \\
\hline
$0.30$ & $-1.115$ & $0.74$  & $\|$ & $1.35$ & $0.82$   \\
\hline\hline
$0.34$ & $-1.03$ & $0.67$  & $\|$ & $1.31$ & $0.52$   \\
\hline
$0.34$ & $-1.03$ & $0.70$  & $\|$ & $1.26$ & $0.52$  \\
\hline
$0.34$ & $-1.03$ & $0.74$  & $\|$ & $1.19$ & $0.52$  \\
\hline
$0.34$ & $-1.185$ & $0.67$  & $\|$ & $1.49$ & $0.82$ \\
\hline
$0.34$ & $-1.185$ & $0.70$  & $\|$ & $1.43$ & $0.82$   \\
\hline
$0.34$ & $-1.185$ & $0.74$  & $\|$ & $1.35$ & $0.82$ \\
\hline\hline
\end{tabular}
\end{table}
\hspace*{3.5mm}Now let's inspect carefully whether the model satisfies the
constraints from the
astronomical observations and the standard Big Bang model. \\
\hspace*{3mm}$\mathbf{1}. \mathbf{Age}\ \mathbf{of}\ \mathbf{the}\
\mathbf{Universe}$\\
\hspace*{6mm}Table $\textrm{I}$ shows that many combinations
predict the ages of the Universe in agreement with
$12.6^{+3.4}_{-2.4}$ Gyr determined from globular clusters age
[17] and $12.5\pm 3.5$ Gyr from radioisotopes studies [18]. It
also doe not conflict the result of $14.1^{+1.0}_{-0.9}$ Gyr from
WMAP, ADSS and SN Ia data
[15].\\
\hspace*{3mm}$\mathbf{2}. \mathbf{Transition}\
\mathbf{Redshift}\ \mathbf{of}\  \mathbf{the}\  \mathbf{Universe}$\\
\hspace*{6mm}The range of transition redshift from deceleration to
acceleration of the Universe is constrained by $\Lambda$CDM [19]
as $z_T=0.67$, the current most stringent constraints of combined
GRB$+$SN Ia data [20] $z_T=0.73\pm0.09$ and the joint analysis of
SNe+CMB data $z_T = 0.52\sim0.73$ [21]. To satisfy the constraint
of $z_T=0.52\sim 0.82$, it is required that $w_{x0}$, i.e.,
$w_{x0}=-1.055\sim-0.92$ for $\Omega_{m0}=0.26$,
$w_{x0}=-1.115\sim-0.97$ for $\Omega_{m0}=0.30$ and
$w_{x0}=-1.185\sim-1.03$ for $\Omega_{m0}=0.34$. \\
\hspace*{3mm}$\mathbf{3}. \mathbf{Big}\ \mathbf{Bang}\ \mathbf{Nucleosynthesis}$\\
\hspace*{6mm}The energy density of the early Universe can be
approximated as $\rho=g_{\ast}\pi^2T^4/30$, where $T$ denotes the
temperature and $g_{\ast}$ denotes the effective number of degrees
of freedom by taking into account the variety of particles at
higher temperatures. We have $g_{\ast}=10.75$ and $g_{\ast}=3.36$
at the temperatures $T=10$ MeV and $T=0.1$ MeV, respectively [22].
The energy density required by BBN epoch is between $1.57\times
10^{-84}$ ($T=10$ MeV) and $4.92\times 10^{-93}$ ($T=0.1$ MeV).
Then we have approximately the beginning time of BBN
$\tau_{bbn1}\simeq7.45$ msec and the ending time of BBN
$\tau_{bbn2}\simeq133$ sec. So the model doe not conflict the
result
of ${10^{-2}}\sim 10^2$ sec estimated by the BBN theory [22].\\
\begin{figure}
\begin{center}
\includegraphics[width=7.5cm]{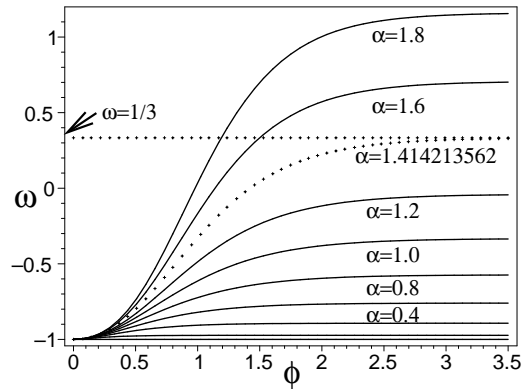}
\end{center}
\caption{$w-\phi$ relations. ($\phi\rightarrow\infty\Longleftrightarrow
z\rightarrow\infty$). Dotted curve is for $\alpha=\sqrt{2}$. When $\alpha=\sqrt{2}$ and
$\phi\rightarrow \infty$, we have $w=1/3$ which is for the radiation-like dominated
Universe. For any $\alpha$, $w\rightarrow-1$ when $\phi\rightarrow 0
\Longleftrightarrow z\rightarrow -1$.} \label{fig:wenvelope}
\end{figure}
\begin{figure}
\begin{center}
\includegraphics[width=8.0cm]{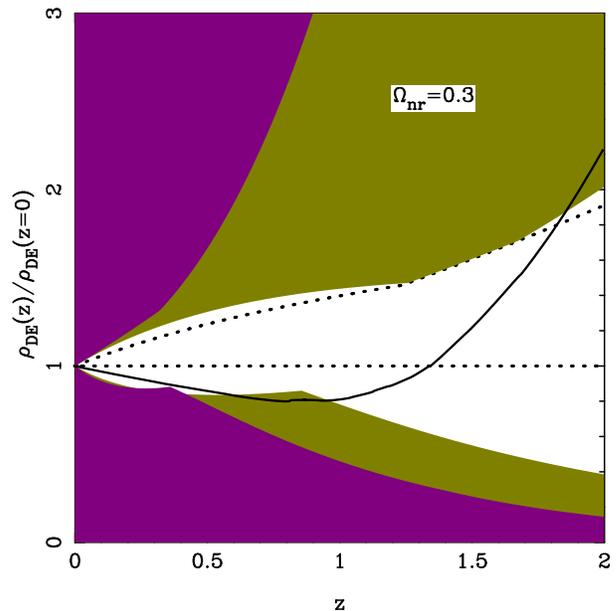}
\end{center}
\caption{Evolutions of the dark energy for $w_{x0}=-1.058$,
over-plotted on the 1-$\sigma$ and 2-$\sigma$ exclusion zones
obtained by combining supernova Ia and WMAP observations [Fig.(2)
(upper panel) in Jassal et al].} \label{fig:wenvelope}
\end{figure}
\hspace*{3mm}$\mathbf{4}.\ \mathbf{Dark}\ \mathbf{Energy}$\\
\hspace*{6mm}When the dilaton field $\phi$ approaches zero, i.e.
the cosmic time $t$ approaches infinity, we have
$\rho=-p=\lambda/(8\pi)$. Then the Universe evolves to de Sitter
Universe in the distant future. So whether the dark energy of the
present Universe is quintessence $w_x>-1$ or phantom $w_x<-1$, it
will evolve into the pure cosmological constant in the future.
Frankly, we can not distinguish the dark energy component from the
total energy density in general. However, since the matter scales
as $\rho_m=\rho_{m0}/a^3$ for low redshift, we can approximate the
dark energy as $\rho_x=\rho-\rho_m$. Because the dark energy was
less important in the past, the most sensitive redshift interval
for probing dark energy is $z =0\sim 2$. Current observations of
SN Ia, CMB and LSS have made constraints [23] on the evolution of
$\rho_x$ at $z=0\sim 2$. FIG.$2$ shows the evolution of the energy
density of dark energy with redshift for one typical parameter in
our models. Our results also satisfy the recent most stringent
constraints which
comes from the results of SN Ia, CMB and LSS [23].\\
\section{conclusion}
In conclusion, we have presented a Big Bang model of the Universe
dominated by dilaton field which can mimic the matter (including
dark matter) and dark energy. The model predicted age of the
Universe, transition redshift, BBN and evolution of dark energy
agree with current observations. Future higher quality data, and
especially from SN Ia data and GRB data (because GRBs are produced
predominantly in the early Universe [24]) at higher redshifts may
constrain the cosmic evolution of matter, dark energy and Hubble
constant more precisely and make critical tests
on our model predictions.\\
\begin{acknowledgments}
\hspace*{3mm} We thank H. K. Jassal, J. S. Bagla and T.
Padmanabhan for sending us their model fitting results we used in
FIG.2. This study is supported in part by the Special Funds for
Major State Basic Research Projects, by the Directional Research
Project of the Chinese Academy of Sciences and by the National
Natural Science Foundation of China. SNZ also acknowledges
supports by NASA's Marshall Space Flight Center and through NASA's
Long Term Space Astrophysics Program.
\end{acknowledgments}

\end{document}